\documentclass[aip,graphicx]{revtex4-1}
\usepackage{svg}
\usepackage{epsfig}
\usepackage{epstopdf}
\usepackage{multirow}

\draft 

\begin{document}


\title{Monitoring Single DNA Docking Site Activity With Sequential Modes of an Optoplasmonic Whispering-Gallery Mode Biosensor} 





 


\author{Narima Eerqing}
\email{narima.eerqing@lshtm.ac.uk}
\affiliation{London School of Hygiene \& Tropical Medicine, London, WC1E 7HT, United Kingdom}
\affiliation{Living Systems Institute, University of Exeter, EX4 4QD}

\author{Ekaterina Zossimova}

\affiliation{Living Systems Institute, University of Exeter, EX4 4QD}
\affiliation{Freiburg Center for Interactive Materials and Bioinspired Technologies, Universität Freiburg}

\author{Sivaraman Subramanian}
\affiliation{Living Systems Institute, University of Exeter, EX4 4QD}

\author{Hsin-Yu Wu}
\affiliation{Living Systems Institute, University of Exeter, EX4 4QD}

\author{Frank Vollmer}
\affiliation{Living Systems Institute, University of Exeter, EX4 4QD}

\date{\today}

\begin{abstract}

    In recent years, there has been rapid advancement in single-molecule techniques, driven by their unparalleled precision in studying molecules whose sizes are beyond the diffraction limit. Among these techniques, optoplasmonic whispering gallery mode sensing has demonstrated great potential in label-free single-molecule characterization. It combines the principles of localized surface plasmon resonance (LSPR) and whispering gallery mode (WGM) sensing, offering exceptional sensing capabilities, even at the level of single ions. However, current optoplasmonic WGM sensing operates in a multiplexed channel, making it challenging to focus on individual binding sites of analyte molecules. In this article, we characterize different binding sites of DNA analyte molecules hybridizing to docking strands on the optoplasmonic WGM sensor, using the ratio of the resonance shift between sequential polar WGM modes. We identify specific docking sites that undergo transient interactions and eventually hybridize with the complementary analyte strands permanently.
    
\end{abstract}


\maketitle 

\section{Introduction}
\label{sec:intro}

The examination of life's processes based on its biomolecular components has been the focus of extensive research spanning several decades. From a physicist's perspective, these processes typically operate well away from thermodynamic equilibrium\cite{fang2019nonequilibrium}. Consequently, predicting the behavior of biomolecules becomes highly challenging due to their existence within complex and non-uniform microbiological systems\cite{Lenn2012,alberts1983studies,nossal2007architecture}. Researchers have employed numerous ensemble averaging techniques to measure the mean state of a biomolecular system. However, even when analyzing a small volume with few biomolecular components, variations can be significant\cite{Miller2017}. Relying solely on bulk measurements to determine an average state can result in an inaccurate estimation of individual molecule states within the system and its processes.

In contrast, single-molecule studies offer an unprecedentedly deep and precise understanding of the biology of biomolecular systems and physics of life\cite{Capitanio2013,Yu2011}. These studies play a pivotal role in examining the multiple metastable free energy states of biomolecules, with energy states separated by multiples of the thermal energy scale\cite{Kufer2009,Hinterdorfer2006}. Importantly, this goes beyond the detection capabilities of conventional bulk measurements. Single-molecule techniques facilitate the observation of specific molecular biological features, contributing to the creation of a comprehensive library of molecular heterogeneity within the system
. Single-molecule fluorescence is one of the most successful techniques for single-molecule detection\cite{weiss1999fluorescence}, relying on the large fluorescence cross-sections of specific organic molecules and proteins. This characteristic enables the detection of fluorescence photons without interference from background signals\cite{Reinhardt2023}. Additionally, single-molecule fluorescence resonance energy transfer (FRET) provides valuable insights into the dynamics of biomolecules and chemicals by leveraging distance-dependent energy transfer between donor and acceptor fluorophores attached to a molecule. Due to the requirement for sufficiently high fluorescence quantum efficiencies, fluorophores (labels) must be attached to target molecules in many single-molecule experiments. However, the label often constrains the natural dynamics of the analyte molecule, since it has a similar molecular weight to the analyte molecule\cite{deniz2008single}. 


On the other hand, localized surface plasmon resonance (LSPR) sensors represent a single-molecule detection approach free of labels. LSPR sensors normally employ noble metals like gold or silver as the plasmonic nanoparticles. These particles can confine light at their hotspot, and molecules interacting with the hotspots can be detected according to the change in the polarizability and refractive index.\cite{baaske2014single,Baaske2022,lin2021click,seth2022high}
Among most of the LSPR-based single-molecule techniques, the optoplasmonic WGM approach offers unique avenues for investigating intricate biomolecular processes that pose challenges for monitoring using alternative single-molecule techniques, including fluorescence-based methods, optical and magnetic tweezers, and atomic force microscopes (AFM). For example, optoplasmonic WGM sensors have showcased the ability to discern subnanometer-scale conformational changes in active enzymes like MalL with microsecond temporal precision. This real-time monitoring facilitates the observation of enzyme conformational states and enables measurements of thermodynamic parameters such as activation heat capacity.

However, even with these cutting-edge single-molecule techniques developed recently, it remains challenging to identify and track reactions that happened on each reaction site. In this article, we provide a novel approach capable of identifying and tracing single DNA docking site activity using an optoplasmonic sensor.

\begin{figure}[tb]
    \centering
    \includegraphics[width=0.8\textwidth]{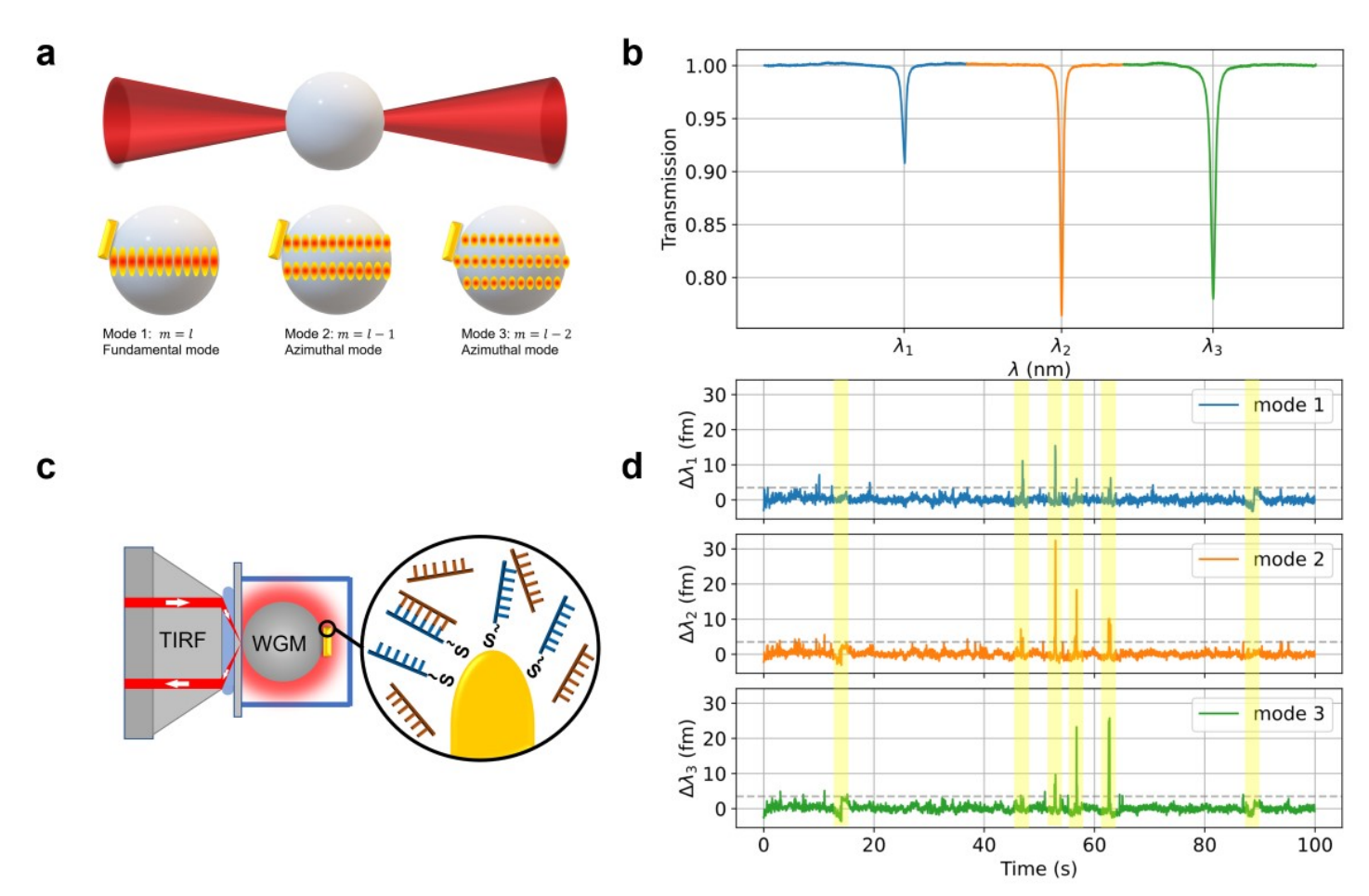}
    \caption{(a)The schematic graph of the whispering gallery polar modes. Gold nanorods are attached to the WGM sensor and overlap with three adjacent modes corresponding to the mode order. (b) Transmission spectrum showing the 3 modes used for the detection of DNA molecules in experiments. (c) Schematic graph of experimental design. An incident laser beam is focused onto the coverslip surface, establishing the total internal reflection. The generated evanescent wave is subsequently coupled into the WGM. The LSPR is excited by the gold nanorods deposited on the surface of the WGM microsphere. (d) Example data-trace of optoplasmonic sensing signals showing the resonance shift, $\Delta \lambda$, experienced by each biosensing mode.}
    \label{fig:1}
\end{figure}

\begin{figure}[tb]
    \centering
    \includegraphics[width=0.8\textwidth]{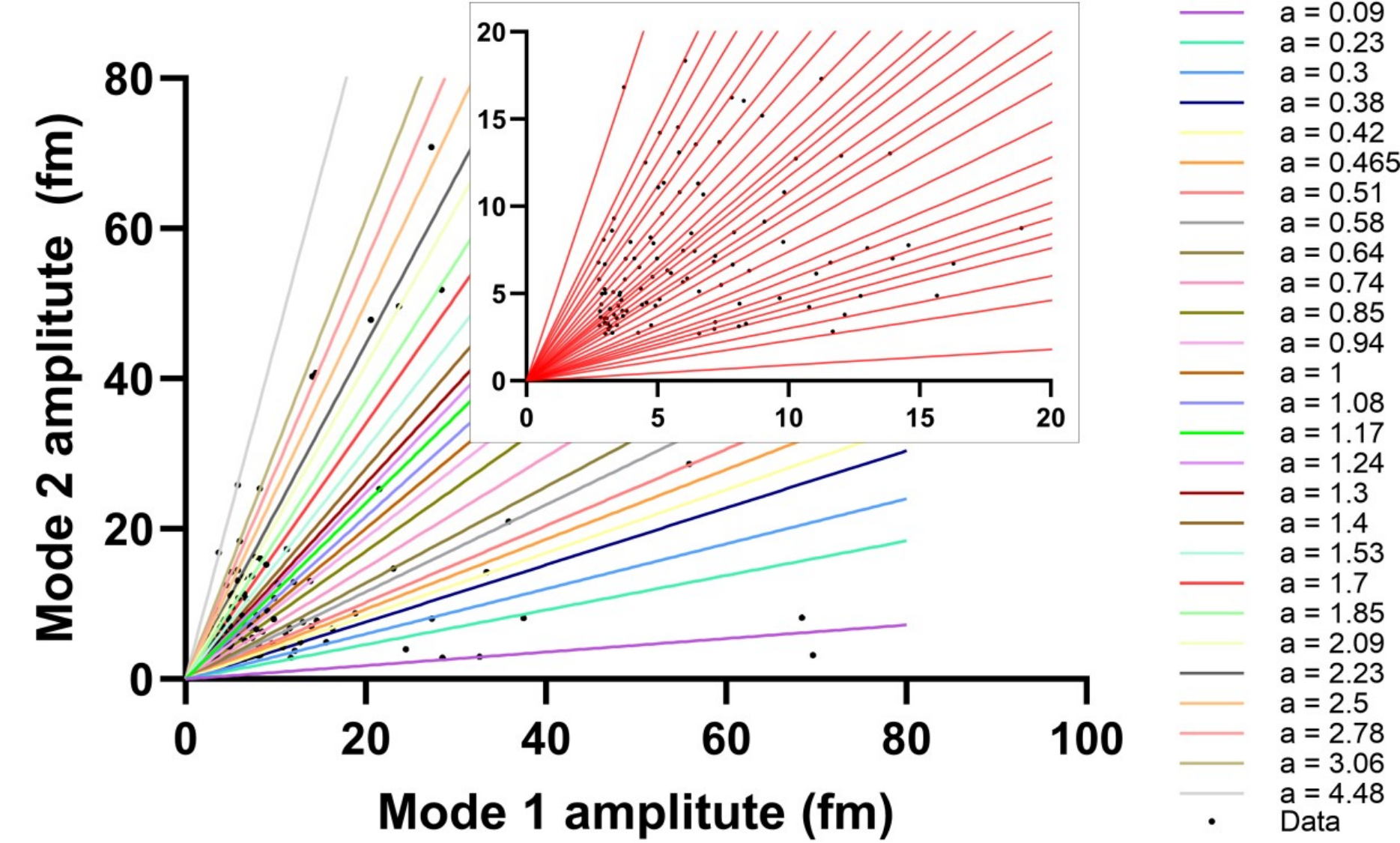}
    \caption{Spike amplitudes of Mode 1 and 2 (black dots) collected from first 45mins of 1 $\mu$M ImT22 hybridization experiment. Linear fitting is carried out for the event number greater than 1 (With bin width=0.02). Corresponding slopes ($a$) are shown and denoted as different colors. Inset figure is the zoomed-in view of Mode 1 and 2 amplitudes within 20fm. Among all collected data, only 12 data points appeared only once, the remaining 115 data points are shown in repeated slopes. }
    \label{fig:2}
\end{figure}

\section{Methods}
\subsection*{Optoplasmonic Sensing of DNA Hybridisation}

The experiments are conducted on the SIMOPS (single-molecule imaging microscopy and optoplasmonic sensing) platform (Figure 1a). TIRF (Total Internal Reflection Fluorescence) microscopy is utilized to excite evanescent waves, while the external cavity laser scans narrowly around the WGM resonance bandwidth to capture the resonance spectra, and an approximately 80 µm diameter glass sphere is employed to generate WGMs. Gold nanorods (GNRs) are irreversibly immobilized on the glass surface, and LSPR is excited through the WGMs. The docking strands (Figure 1a, denoted in blue) are immobilized on the GNRs via a mercaptohexyl linker. Figure 1b illustrates the transmission spectrum of the three sequential modes used in the experiments. Figure 1c depicts the corresponding WGM mode numbers, where the first mode from the left (Figure 1b, blue) represents the fundamental mode, and the others are higher-order modes.

As complementary strands (imagers) approach the docking strands located in the vicinity of the GNRs hotspots, shifts occur in the WGM resonance due to localized alterations in the refractive index (essentially induced polarization of the molecule by the electric field, leading to a wavelength shift $\Delta\lambda$). Figure 1d displays an example wavelength change time trace of optoplasmonic sensing from three sequential modes. A centroid fitting algorithm is employed to enable extraction of $\Delta\lambda$ and $\Delta\kappa$ from the resonance spectra of multiple WGM modes. Custom MATLAB code is utilized for peak detection, background trend removal (detrending), and signal height measurement. Spike signal detection is performed using a 3$\sigma$ threshold. \cite{baaske2014single}

\subsection*{Sequential polar modes of an optoplasmonic resonator}

The single-molecule sensing platform comprises a whispering gallery mode (WGM) and several gold nanorods (GNRs). This methodology involves the excitation of two resonances characterized by the same angular momentum quantum number ($\ell$) but differing magnetic quantum numbers ($m$) within the range of ($-\ell < m < \ell$) in a cavity. While $m$ is traditionally known as the magnetic quantum number in atomic physics, when analyzing various $m$ modes within a microresonator with a given $\ell$, the term "polar modes" will be employed. To comprehend the intricacies of this approach, it is essential to scrutinize the polar Whispering Gallery Mode (WGM) intensity distribution.

\begin{figure}[tb]
    \centering
    \includegraphics[width=0.9\textwidth]{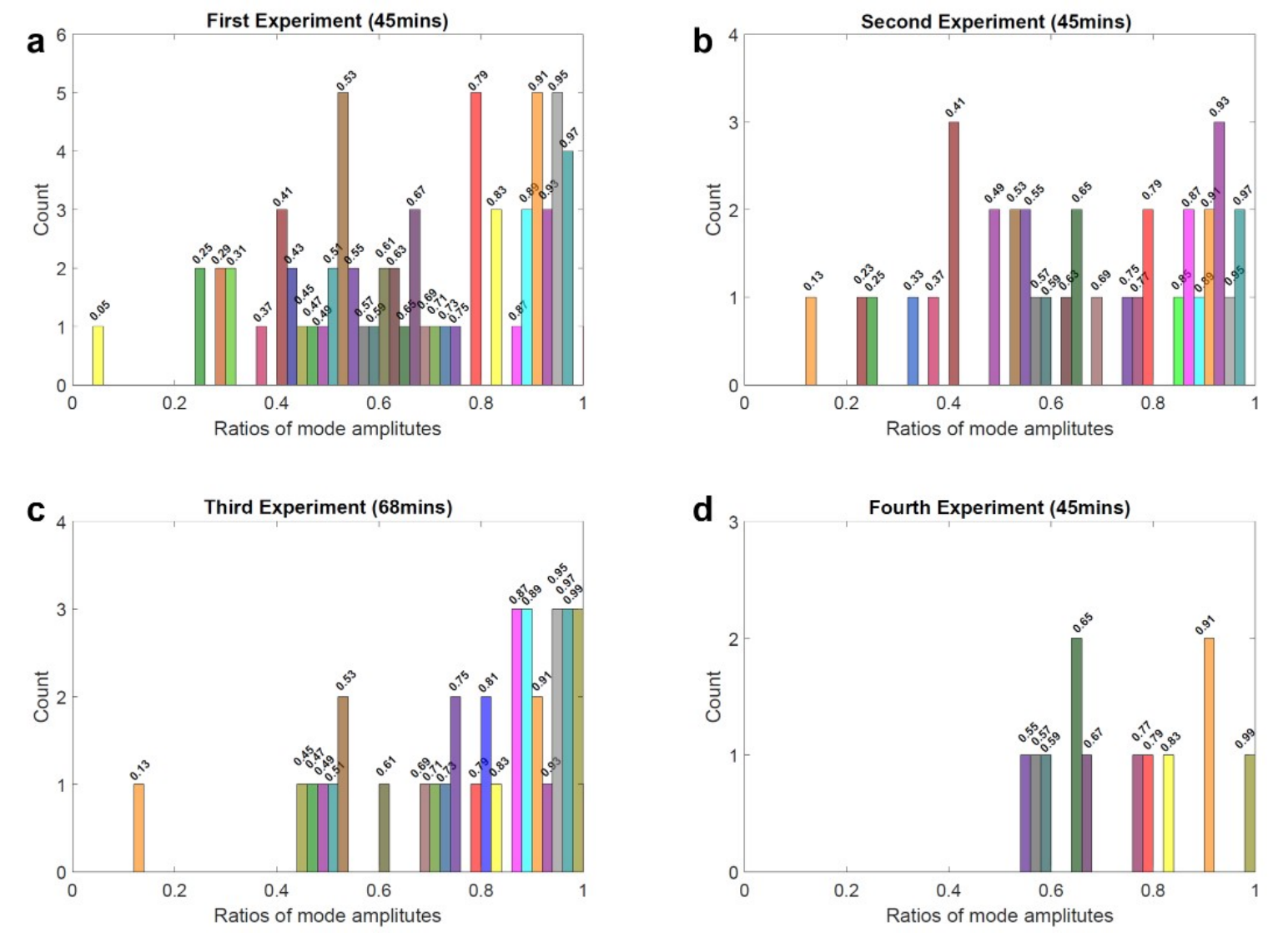}
    \caption{Part of spike amplitude ratios (Ratios between 0 and 1) histogram of Mode3 and Mode2 collected from four continuous optoplasmonic sensing experiments (Bin width=0.02). The ratios among different experiments are labelled with specific colors.}
    \label{fig:3}
\end{figure}

The number of intensity peaks along the polar direction is given by $\ell-m+1$. In a spherical configuration, these $m$ states exhibit degeneracy for a specific angular momentum $\ell$. However, in geometries where the perfect spherical symmetry is broken, such as a spheroid, this degeneracy is resolved, leading to spectral separation of the states. The fundamental mode is an equatorial mode with $m=\ell$, resulting in a single intensity peak centred around the equator. Subsequently, the $m=\ell-1$ mode displays two peaks, positioned to the North and South of the equator (see Figure 1c). It is crucial to note that these three modes depicted in Figure 1b are activated sequentially within the same microcavity. Similar experiments employing sequential polar modes have been shown by D. Keng et al\cite{Keng2014}. They used fibre-coupled WGM to detect nanoparticles as they absorb on the glass sphere; ratios of sequential modes are employed to determine the latitude angle of where the particles land. In the SIMOPS system, the GNRs are fixed on the WGM microsphere, and single-stranded DNAs are perturbing the sensor. According to the study by D. Keng et al., the number of shift ratios should be identical to the number of GNR perturbations. However, we observe many more shift ratios exceeding the total number of GNRs, which indicates that other factors exist causing the various ratios. 

There are approximately 10 gold nanorods in the experiment, but not all of the nanorods will bind at the equator or be oriented at the most optimal angle for maximum field enhancement. It is difficult to control where the gold nanorods will bind in experiments. The plasmon resonance of the gold nanorod is very sensitive to the polarization direction of the incident light wave. The maximum field enhancement occurs when the long axis of the nanorod is parallel to the polarization direction of the incident wave. Since the nanorods are scattered randomly across the WGM surface, each plasmonic hot spot will have a unique electric field intensity distribution. In addition, it is well known that surface roughness dramatically affects field enhancement for plasmonic nanoparticles\cite{baaske2014single,novotny2012principles}. The docking sites are randomly distributed on the GNRs, the biosensor signal depends on the intensity of the field at the position of the molecule, therefore it is highly likely that this uneven surface would lead to heterogeneity in the local field and further result in different ratios for DNA molecules binding to GNR docking sites.

\begin{figure}[tb]
    \centering
    \includegraphics[width=0.6\textwidth]{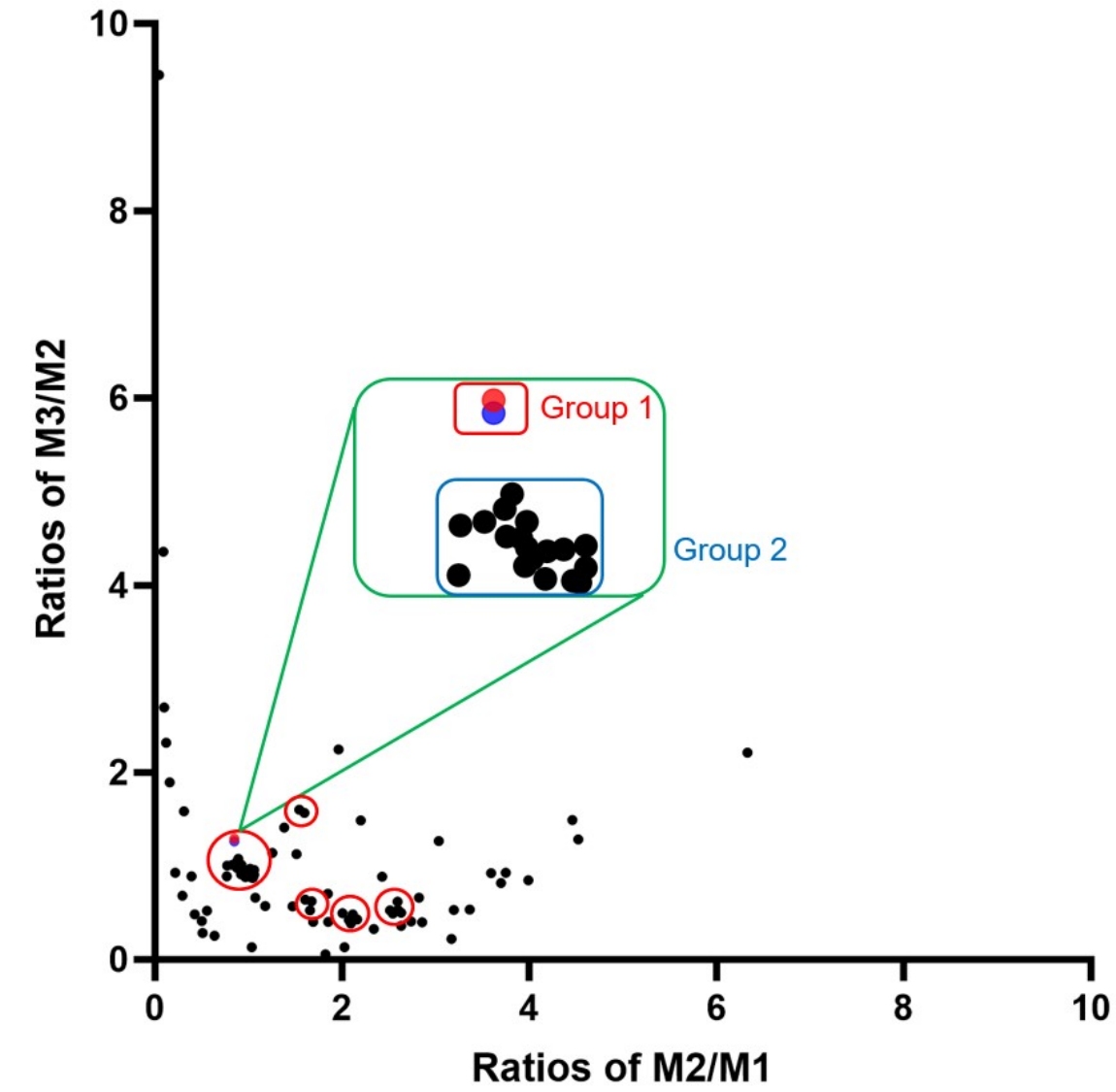}
    \caption{Overall RRAs, RRAs with similar values form clusters. The red boxes (Group 1) in the inset indicate points likely from the same docking sites, while the blue box (Group 2) shows a cluster that may represent one or more docking sites. }
    \label{fig:4}
\end{figure}

\begin{table}[hb]
\renewcommand{\arraystretch}{1.5}
\centering
 \caption{Sequences of ssDNA used for the experiments.}
\label{tab:T1}
\begin{tabular}{lll}
\hline
& \textbf{ssDNA}   & \textbf{Sequence (5'-3') }\\
\hline
\multirow{3}{4em}{Set I} & P1               & [ThiolC6] TTT T\underline{AT ACA TCT A}          \\ 
                         & ImP1*D           & [DY782] C\underline{TA GAT GTA T}                 \\
    
\hline
\multirow{3}{4em}{Set II}& T22              & [ThiolC6] TTT T\underline{GA GAT AAA} C\underline{GA} G\underline{AA GGA} T\underline{TG AT} \\
                 & ImT22*D           & [DY782] \underline{ATC A}G\underline{T CCT T}T\underline{T C}C\underline{T TTA TCT C}  (3 mismatched)  \\
\hline
\end{tabular}
\end{table}

\section{Results}
\subsection*{Repeating amplitude ratios among adjacent modes}

Spike signal amplitudes for single molecule DNA experiments are collected from a 45-minute optoplasmonic sensing experiment with hybridization between 1$\mu$M of ImT22 imager strands (Sequences are shown in Table 1.) and complementary docking sites on GNRs. We plot the amplitudes for events that give a signal above the noise threshold for Mode 1 versus Mode 2 in Figure 2. 

Linear fitting was applied to event numbers greater than 1. The corresponding slopes (denoted as 'a') are displayed in various colors. The inset figure provides a zoomed view of the Mode 1 and Mode 2 amplitudes within the range of 20 fm. Among the entire dataset, only 12 data points only shown once, while the remaining 115 data points can be fitted well into a linear function $y=a \cdot x$, indicating that repeating ratios of WGM mode amplitudes (RRA) exist among adjacent polar modes. Further analysis is carried out for $m=\ell-1$ (Mode 2) and $m=\ell-2$ (Mode 3) for four continuous experiments, and the observation is consistent with Mode 1 and Mode 2 results. Part of the ratios are plotted as the histogram graph in Figure 3; only ratios between 0 and 1 are plotted, as the ratios greater than this also follow the same patterns but they are distributed sparsely and therefore neglected. 

Two main characters can be concluded from these four subfigures. Firstly, many RRAs exist across different experiments, some of which show more frequently than others, most likely due to the corresponding docking sites being more accessible. While there are also rare ratios only shown once or twice during experiments over 3 hours, suggesting a less accessible docking site. Secondly, there is a decreasing trend in the number of events. The first 45 minutes of the experiment contain the highest number of events, most ratios gradually disappear over time. In the fourth experiment, only a few ratios remain. This observation matches with what we have reported \cite{eerqing2023anomalous} before; we attributed the reason to the anomalous permanent hybridization between the docking sites and imager strands.

\subsection*{Tracing of individual docking site interactions}

Thus far, we have demonstrated RRAs among the adjacent polar modes (either Mode 2 $\&$ Mode 1 or Mode 3 $\&$ Mode 2). These ratios illustrate how adjacent polar modes view the same event. One may argue that if these ratios are close to each other, it would be challenging to distinguish them. Here, we present a simple approach to identifying individual sites. If an event is simultaneously captured by all three polar modes, one can obtain two ratios for it: Mode 2 $\&$ Mode 1 and Mode 3 $\&$ Mode 2. These two ratios are akin to two projections of the same object from different angles. While some objects may appear similar when viewed from one angle alone, they are likely to be different when observed from another perspective.

To label the ratios, we have used ratio identifiers (RRAs), which are composed of two numbers. The first one represents the ratio of Mode 2 over Mode 1 amplitudes, and the second number is the ratio of Mode 3 over Mode 2 amplitudes. Figure 4 displays all RRAs from the four experiments, showing several clusters of data points, indicating that those events are likely from the same reaction site. We can easily observe that even though some RRAs share the same ratios between two adjacent modes, they are actually from different sites. We have demonstrated the ability to distinguish single docking sites from multiplexed channels, making it possible to trace the activity from individual docking sites.

\begin{figure}[tb]
    \centering
    \includegraphics[width=0.8\textwidth]{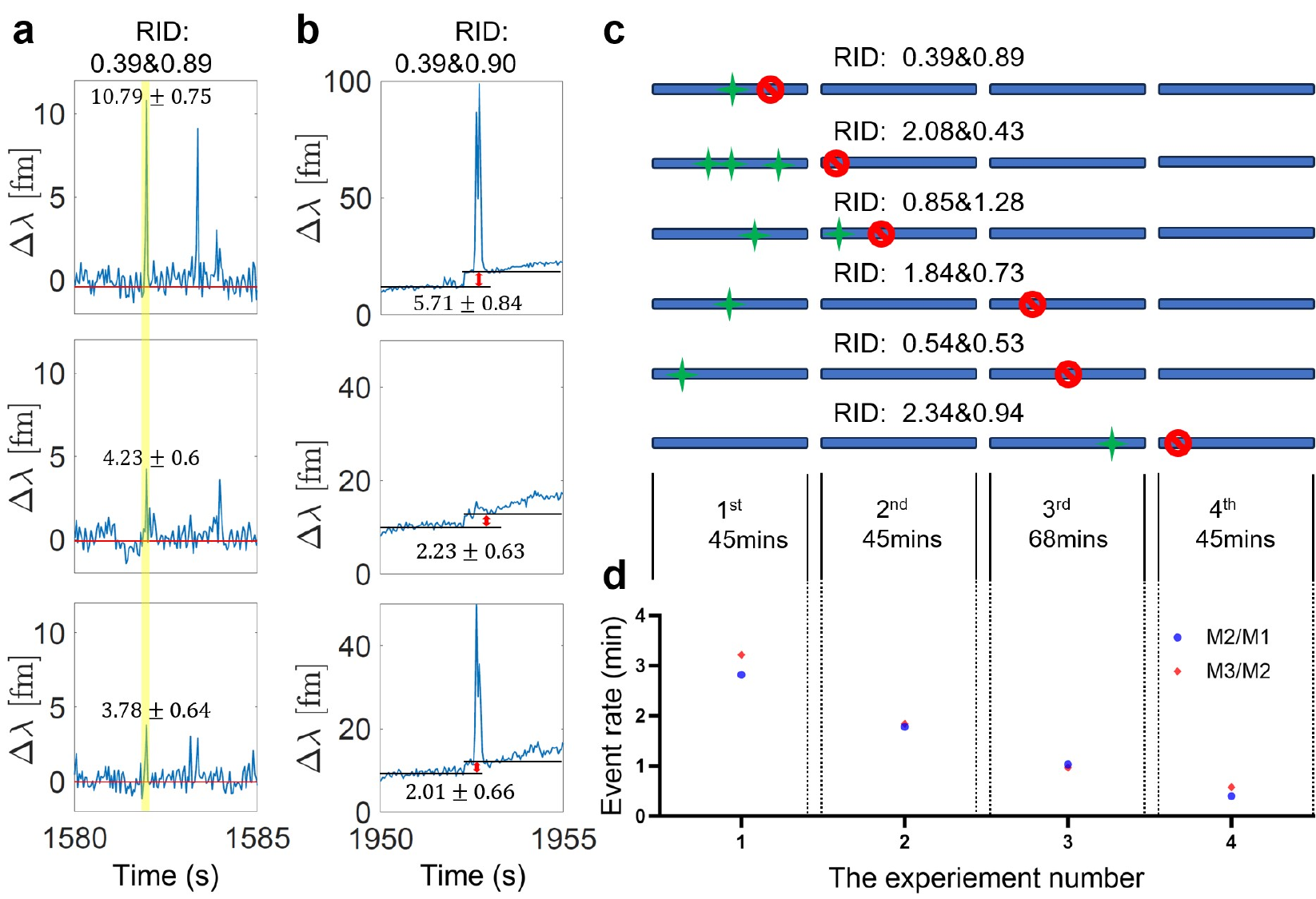}
    \caption{ (a)Example of a transient interaction (Highlighted event) captured by sequential polar modes and their amplitude. The ratio identifier (RRA) denotes the corresponding amplitude ratios between (Mode2/Mode1 \& Mode3/Mode2). (b) The step signal with the same RRA is detected after the spike event. (c) Spike (Green star) and step (Red stop sign) signals with the same RRAs captured within four continuous experiments. No spike signals are observed after a step signal, indicating the site is no longer accessible. (d) Event rates per minute for the four continuous experiments. Events captured by both Mode 1 and 2 are denoted as blue dots, by both Mode 2 and 3 are denoted as red dots.}
    \label{fig:5}
\end{figure}

\subsection*{Step signals correspond to the disappeared ratios}

For the optoplasmonic sensor, transient interactions are captured in the form of spike signals, where the molecules enter the hotspot of plasmonic nanoparticles and leave shortly after. However, when a binding event occurs, the molecule interacts and stays with the sensor. The polarizability and refractive index within the hotspot are changed, causing a red shift to the WGM spectrum, and therefore observed as a step signal. If the missing ratios correspond to anomalous hybridizations, it is expected that the ratios of the step heights correlate with the ratios of missing spike signals, RRA of spike signal is no longer observed after RRA of step signal indicating permanent hybridisation at single docking site.

We then compared the ratios extracted from the spike signals that showed in the early experiment but then disappeared in the latter experiments with the step signals, in this process, we noticed that some disappeared ratios match with step ratios. We compare RRA step signals to RRA spike signals that occur before the step signal and confirm that these RRAs no longer occur after the step signal.
An example of a spike signal (highlighted in yellow) during the first 45-minute experiment is shown in Figure 5a; it is detected by three polar modes. Using the signal amplitude, we can calculate the RRA, which equals $0.39\&0.89$. A step signal is found after around 370 seconds (Figure 5b); using the same method, we can calculate the RRA, which equals $0.39\&0.90$. We then checked the following three experiments, and no spikes or steps with similar RRA were found.

Similar characterization is conducted and shown in Figure 5c; the green stars denote the time when the spike signal occurs, and the red stop sign corresponds to the time when the step signal appears. This result further supports our hypothesis of anomalous hybridization, suggesting that the reduced event rates are caused by the permanent hybridizations between the docking strands and the complementary strands, which blocks the docking sites permanently.

\section{discussion}

From the wavelength shift time trace, we often observe a shift detected by one or two modes, but no signal is detected by the other mode(s). This is most likely due to the corresponding signal from the other mode(s) falling below the $3\sigma$ noise threshold, and only a few modes can be detected by all three modes. This limitation has restricted the capability of employing more WGMs\cite{vollmer2008whispering}, since it is expected that even fewer events can be detected with more modes simultaneously. Therefore, a new method to increase the overall signal-to-noise level is desired to improve data acquisition efficiency.

We have demonstrated the repeating ratios between adjacent polar modes and the approach to identifying and tracing docking sites using the ratios of the resonance shift between sequential polar modes. We have used this method to trace individual docking sites that undergo transient interactions and eventually bond to the docking strands permanently. Evidence shows that the blocked docking sites have lost all of their accessibility, and no further interaction is detected thereafter, matching the observations from previous studies \cite{eerqing2023anomalous}.

The RRA approach could become a powerful method to separate signals originating from different sites on many other multiplexed sensing techniques like SPR sensing, and so on, apart from hybridisation these could be signals originating from enzyme activity, or nanoscale catalysis\cite{houghton2024single}. It may also be possible to purposely immobilise different docking strands on the sensor and identify the RRA corresponding to a specific sequence in sequential single molecule experiments similar in approach. By using a DNA origami approach\cite{rothemund2006folding}, for example origami with a nanorod dimer and several different docking sites in the dimer hotspot it may be possible to show that the sensor is well capable of resolving signals from docking sites separated by Angstrom resolution. Taking this idea further, perhaps the sensing signals from proteins labelled with DNA nanobodies allow us to obtain some information on the composition and orientation of a protein aggregate structure on the sensor\cite{jager2006site}.


\section{Acknowledgement}

The authors would like to express sincere gratitude to Professor Jonathan Pines for his invaluable support and guidance during the time at the Institute of Cancer Research. His mentorship and encouragement were instrumental throughout the course of this work.

%


\bibliography{biosensor}

\begin{thebibliography}{23}%
\makeatletter
\providecommand \@ifxundefined [1]{%
 \@ifx{#1\undefined}
}%
\providecommand \@ifnum [1]{%
 \ifnum #1\expandafter \@firstoftwo
 \else \expandafter \@secondoftwo
 \fi
}%
\providecommand \@ifx [1]{%
 \ifx #1\expandafter \@firstoftwo
 \else \expandafter \@secondoftwo
 \fi
}%
\providecommand \natexlab [1]{#1}%
\providecommand \enquote  [1]{``#1''}%
\providecommand \bibnamefont  [1]{#1}%
\providecommand \bibfnamefont [1]{#1}%
\providecommand \citenamefont [1]{#1}%
\providecommand \href@noop [0]{\@secondoftwo}%
\providecommand \href [0]{\begingroup \@sanitize@url \@href}%
\providecommand \@href[1]{\@@startlink{#1}\@@href}%
\providecommand \@@href[1]{\endgroup#1\@@endlink}%
\providecommand \@sanitize@url [0]{\catcode `\\12\catcode `\$12\catcode
  `\&12\catcode `\#12\catcode `\^12\catcode `\_12\catcode `\%12\relax}%
\providecommand \@@startlink[1]{}%
\providecommand \@@endlink[0]{}%
\providecommand \url  [0]{\begingroup\@sanitize@url \@url }%
\providecommand \@url [1]{\endgroup\@href {#1}{\urlprefix }}%
\providecommand \urlprefix  [0]{URL }%
\providecommand \Eprint [0]{\href }%
\providecommand \doibase [0]{http://dx.doi.org/}%
\providecommand \selectlanguage [0]{\@gobble}%
\providecommand \bibinfo  [0]{\@secondoftwo}%
\providecommand \bibfield  [0]{\@secondoftwo}%
\providecommand \translation [1]{[#1]}%
\providecommand \BibitemOpen [0]{}%
\providecommand \bibitemStop [0]{}%
\providecommand \bibitemNoStop [0]{.\EOS\space}%
\providecommand \EOS [0]{\spacefactor3000\relax}%
\providecommand \BibitemShut  [1]{\csname bibitem#1\endcsname}%
\let\auto@bib@innerbib\@empty
\bibitem [{\citenamefont {Fang}\ \emph {et~al.}(2019)\citenamefont {Fang},
  \citenamefont {Kruse}, \citenamefont {Lu},\ and\ \citenamefont
  {Wang}}]{fang2019nonequilibrium}%
  \BibitemOpen
  \bibfield  {author} {\bibinfo {author} {\bibfnamefont {X.}~\bibnamefont
  {Fang}}, \bibinfo {author} {\bibfnamefont {K.}~\bibnamefont {Kruse}},
  \bibinfo {author} {\bibfnamefont {T.}~\bibnamefont {Lu}}, \ and\ \bibinfo
  {author} {\bibfnamefont {J.}~\bibnamefont {Wang}},\ }\bibfield  {title}
  {\enquote {\bibinfo {title} {Nonequilibrium physics in biology},}\
  }\href@noop {} {\bibfield  {journal} {\bibinfo  {journal} {Reviews of Modern
  Physics}\ }\textbf {\bibinfo {volume} {91}},\ \bibinfo {pages} {045004}
  (\bibinfo {year} {2019})}\BibitemShut {NoStop}%
\bibitem [{\citenamefont {Lenn}\ and\ \citenamefont {Leake}(2012)}]{Lenn2012}%
  \BibitemOpen
  \bibfield  {author} {\bibinfo {author} {\bibfnamefont {T.}~\bibnamefont
  {Lenn}}\ and\ \bibinfo {author} {\bibfnamefont {M.~C.}\ \bibnamefont
  {Leake}},\ }\bibfield  {title} {\enquote {\bibinfo {title} {{Experimental
  approaches for addressing fundamental biological questions in living,
  functioning cells with single molecule precision}},}\ }\href@noop {}
  {\bibfield  {journal} {\bibinfo  {journal} {Open biology}\ }\textbf {\bibinfo
  {volume} {2}},\ \bibinfo {pages} {120090} (\bibinfo {year}
  {2012})}\BibitemShut {NoStop}%
\bibitem [{\citenamefont {Alberts}\ \emph {et~al.}(1983)\citenamefont
  {Alberts}, \citenamefont {Barry}, \citenamefont {Bedinger}, \citenamefont
  {Formosa}, \citenamefont {Jongeneel},\ and\ \citenamefont
  {Kreuzer}}]{alberts1983studies}%
  \BibitemOpen
  \bibfield  {author} {\bibinfo {author} {\bibfnamefont {B.}~\bibnamefont
  {Alberts}}, \bibinfo {author} {\bibfnamefont {J.}~\bibnamefont {Barry}},
  \bibinfo {author} {\bibfnamefont {P.}~\bibnamefont {Bedinger}}, \bibinfo
  {author} {\bibfnamefont {T.}~\bibnamefont {Formosa}}, \bibinfo {author}
  {\bibfnamefont {C.}~\bibnamefont {Jongeneel}}, \ and\ \bibinfo {author}
  {\bibfnamefont {K.}~\bibnamefont {Kreuzer}},\ }\bibfield  {title} {\enquote
  {\bibinfo {title} {Studies on dna replication in the bacteriophage t4 a--.
  gif system},}\ }in\ \href@noop {} {\emph {\bibinfo {booktitle} {Cold Spring
  Harbor symposia on quantitative biology}}},\ Vol.~\bibinfo {volume} {47}\
  (\bibinfo {organization} {Cold Spring Harbor Laboratory Press},\ \bibinfo
  {year} {1983})\ pp.\ \bibinfo {pages} {655--668}\BibitemShut {NoStop}%
\bibitem [{\citenamefont {Nossal}\ \emph {et~al.}(2007)\citenamefont {Nossal},
  \citenamefont {Makhov}, \citenamefont {Chastain}, \citenamefont {Jones},\
  and\ \citenamefont {Griffith}}]{nossal2007architecture}%
  \BibitemOpen
  \bibfield  {author} {\bibinfo {author} {\bibfnamefont {N.~G.}\ \bibnamefont
  {Nossal}}, \bibinfo {author} {\bibfnamefont {A.~M.}\ \bibnamefont {Makhov}},
  \bibinfo {author} {\bibfnamefont {P.~D.}\ \bibnamefont {Chastain}}, \bibinfo
  {author} {\bibfnamefont {C.~E.}\ \bibnamefont {Jones}}, \ and\ \bibinfo
  {author} {\bibfnamefont {J.~D.}\ \bibnamefont {Griffith}},\ }\bibfield
  {title} {\enquote {\bibinfo {title} {Architecture of the bacteriophage t4
  replication complex revealed with nanoscale biopointers},}\ }\href@noop {}
  {\bibfield  {journal} {\bibinfo  {journal} {Journal of Biological Chemistry}\
  }\textbf {\bibinfo {volume} {282}},\ \bibinfo {pages} {1098--1108} (\bibinfo
  {year} {2007})}\BibitemShut {NoStop}%
\bibitem [{\citenamefont {Miller}\ \emph {et~al.}(2017)\citenamefont {Miller},
  \citenamefont {Zhou}, \citenamefont {Shepherd}, \citenamefont {Wollman},\
  and\ \citenamefont {Leake}}]{Miller2017}%
  \BibitemOpen
  \bibfield  {author} {\bibinfo {author} {\bibfnamefont {H.}~\bibnamefont
  {Miller}}, \bibinfo {author} {\bibfnamefont {Z.}~\bibnamefont {Zhou}},
  \bibinfo {author} {\bibfnamefont {J.}~\bibnamefont {Shepherd}}, \bibinfo
  {author} {\bibfnamefont {A.~J.~M.}\ \bibnamefont {Wollman}}, \ and\ \bibinfo
  {author} {\bibfnamefont {M.~C.}\ \bibnamefont {Leake}},\ }\bibfield  {title}
  {\enquote {\bibinfo {title} {{Single-molecule techniques in biophysics: a
  review of the progress in methods and applications}},}\ }\href@noop {}
  {\bibfield  {journal} {\bibinfo  {journal} {Reports on Progress in Physics}\
  }\textbf {\bibinfo {volume} {81}},\ \bibinfo {pages} {24601} (\bibinfo {year}
  {2017})}\BibitemShut {NoStop}%
\bibitem [{\citenamefont {Capitanio}\ and\ \citenamefont
  {Pavone}(2013)}]{Capitanio2013}%
  \BibitemOpen
  \bibfield  {author} {\bibinfo {author} {\bibfnamefont {M.}~\bibnamefont
  {Capitanio}}\ and\ \bibinfo {author} {\bibfnamefont {F.~S.}\ \bibnamefont
  {Pavone}},\ }\bibfield  {title} {\enquote {\bibinfo {title} {{Interrogating
  biology with force: single molecule high-resolution measurements with optical
  tweezers}},}\ }\href@noop {} {\bibfield  {journal} {\bibinfo  {journal}
  {Biophysical journal}\ }\textbf {\bibinfo {volume} {105}},\ \bibinfo {pages}
  {1293--1303} (\bibinfo {year} {2013})}\BibitemShut {NoStop}%
\bibitem [{\citenamefont {Yu}\ \emph {et~al.}(2011)\citenamefont {Yu},
  \citenamefont {Chen}, \citenamefont {Qu},\ and\ \citenamefont
  {Niu}}]{Yu2011}%
  \BibitemOpen
  \bibfield  {author} {\bibinfo {author} {\bibfnamefont {B.}~\bibnamefont
  {Yu}}, \bibinfo {author} {\bibfnamefont {D.}~\bibnamefont {Chen}}, \bibinfo
  {author} {\bibfnamefont {J.}~\bibnamefont {Qu}}, \ and\ \bibinfo {author}
  {\bibfnamefont {H.}~\bibnamefont {Niu}},\ }\bibfield  {title} {\enquote
  {\bibinfo {title} {{Fast Fourier domain localization algorithm of a single
  molecule with nanometer precision}},}\ }\href@noop {} {\bibfield  {journal}
  {\bibinfo  {journal} {Optics letters}\ }\textbf {\bibinfo {volume} {36}},\
  \bibinfo {pages} {4317--4319} (\bibinfo {year} {2011})}\BibitemShut {NoStop}%
\bibitem [{\citenamefont {Kufer}\ \emph {et~al.}(2009)\citenamefont {Kufer},
  \citenamefont {Strackharn}, \citenamefont {Stahl}, \citenamefont {Gumpp},
  \citenamefont {Puchner},\ and\ \citenamefont {Gaub}}]{Kufer2009}%
  \BibitemOpen
  \bibfield  {author} {\bibinfo {author} {\bibfnamefont {S.~K.}\ \bibnamefont
  {Kufer}}, \bibinfo {author} {\bibfnamefont {M.}~\bibnamefont {Strackharn}},
  \bibinfo {author} {\bibfnamefont {S.~W.}\ \bibnamefont {Stahl}}, \bibinfo
  {author} {\bibfnamefont {H.}~\bibnamefont {Gumpp}}, \bibinfo {author}
  {\bibfnamefont {E.~M.}\ \bibnamefont {Puchner}}, \ and\ \bibinfo {author}
  {\bibfnamefont {H.~E.}\ \bibnamefont {Gaub}},\ }\bibfield  {title} {\enquote
  {\bibinfo {title} {{Optically monitoring the mechanical assembly of single
  molecules}},}\ }\href@noop {} {\bibfield  {journal} {\bibinfo  {journal}
  {Nature nanotechnology}\ }\textbf {\bibinfo {volume} {4}},\ \bibinfo {pages}
  {45--49} (\bibinfo {year} {2009})}\BibitemShut {NoStop}%
\bibitem [{\citenamefont {Hinterdorfer}\ and\ \citenamefont
  {Dufr{\^{e}}ne}(2006)}]{Hinterdorfer2006}%
  \BibitemOpen
  \bibfield  {author} {\bibinfo {author} {\bibfnamefont {P.}~\bibnamefont
  {Hinterdorfer}}\ and\ \bibinfo {author} {\bibfnamefont {Y.~F.}\ \bibnamefont
  {Dufr{\^{e}}ne}},\ }\bibfield  {title} {\enquote {\bibinfo {title}
  {{Detection and localization of single molecular recognition events using
  atomic force microscopy}},}\ }\href@noop {} {\bibfield  {journal} {\bibinfo
  {journal} {Nature methods}\ }\textbf {\bibinfo {volume} {3}},\ \bibinfo
  {pages} {347--355} (\bibinfo {year} {2006})}\BibitemShut {NoStop}%
\bibitem [{\citenamefont {Weiss}(1999)}]{weiss1999fluorescence}%
  \BibitemOpen
  \bibfield  {author} {\bibinfo {author} {\bibfnamefont {S.}~\bibnamefont
  {Weiss}},\ }\bibfield  {title} {\enquote {\bibinfo {title} {Fluorescence
  spectroscopy of single biomolecules},}\ }\href@noop {} {\bibfield  {journal}
  {\bibinfo  {journal} {Science}\ }\textbf {\bibinfo {volume} {283}},\ \bibinfo
  {pages} {1676--1683} (\bibinfo {year} {1999})}\BibitemShut {NoStop}%
\bibitem [{\citenamefont {Reinhardt}\ \emph {et~al.}(2023)\citenamefont
  {Reinhardt}, \citenamefont {Masullo}, \citenamefont {Baudrexel},
  \citenamefont {Steen}, \citenamefont {Kowalewski}, \citenamefont {Eklund},
  \citenamefont {Strauss}, \citenamefont {Unterauer}, \citenamefont
  {Schlichthaerle}, \citenamefont {Strauss}, \citenamefont {Klein},\ and\
  \citenamefont {Jungmann}}]{Reinhardt2023}%
  \BibitemOpen
  \bibfield  {author} {\bibinfo {author} {\bibfnamefont {S.~C.~M.}\
  \bibnamefont {Reinhardt}}, \bibinfo {author} {\bibfnamefont {L.~A.}\
  \bibnamefont {Masullo}}, \bibinfo {author} {\bibfnamefont {I.}~\bibnamefont
  {Baudrexel}}, \bibinfo {author} {\bibfnamefont {P.~R.}\ \bibnamefont
  {Steen}}, \bibinfo {author} {\bibfnamefont {R.}~\bibnamefont {Kowalewski}},
  \bibinfo {author} {\bibfnamefont {A.~S.}\ \bibnamefont {Eklund}}, \bibinfo
  {author} {\bibfnamefont {S.}~\bibnamefont {Strauss}}, \bibinfo {author}
  {\bibfnamefont {E.~M.}\ \bibnamefont {Unterauer}}, \bibinfo {author}
  {\bibfnamefont {T.}~\bibnamefont {Schlichthaerle}}, \bibinfo {author}
  {\bibfnamefont {M.~T.}\ \bibnamefont {Strauss}}, \bibinfo {author}
  {\bibfnamefont {C.}~\bibnamefont {Klein}}, \ and\ \bibinfo {author}
  {\bibfnamefont {R.}~\bibnamefont {Jungmann}},\ }\bibfield  {title} {\enquote
  {\bibinfo {title} {Ångstr\"{o}m-resolution fluorescence microscopy},}\
  }\href {\doibase 10.1038/s41586-023-05925-9} {\bibfield  {journal} {\bibinfo
  {journal} {Nature}\ }\textbf {\bibinfo {volume} {617}},\ \bibinfo {pages}
  {711–716} (\bibinfo {year} {2023})}\BibitemShut {NoStop}%
\bibitem [{\citenamefont {Deniz}, \citenamefont {Mukhopadhyay},\ and\
  \citenamefont {Lemke}(2008)}]{deniz2008single}%
  \BibitemOpen
  \bibfield  {author} {\bibinfo {author} {\bibfnamefont {A.~A.}\ \bibnamefont
  {Deniz}}, \bibinfo {author} {\bibfnamefont {S.}~\bibnamefont {Mukhopadhyay}},
  \ and\ \bibinfo {author} {\bibfnamefont {E.~A.}\ \bibnamefont {Lemke}},\
  }\bibfield  {title} {\enquote {\bibinfo {title} {Single-molecule biophysics:
  at the interface of biology, physics and chemistry},}\ }\href@noop {}
  {\bibfield  {journal} {\bibinfo  {journal} {Journal of the Royal Society
  Interface}\ }\textbf {\bibinfo {volume} {5}},\ \bibinfo {pages} {15--45}
  (\bibinfo {year} {2008})}\BibitemShut {NoStop}%
\bibitem [{\citenamefont {Baaske}, \citenamefont {Foreman},\ and\ \citenamefont
  {Vollmer}(2014)}]{baaske2014single}%
  \BibitemOpen
  \bibfield  {author} {\bibinfo {author} {\bibfnamefont {M.~D.}\ \bibnamefont
  {Baaske}}, \bibinfo {author} {\bibfnamefont {M.~R.}\ \bibnamefont {Foreman}},
  \ and\ \bibinfo {author} {\bibfnamefont {F.}~\bibnamefont {Vollmer}},\
  }\bibfield  {title} {\enquote {\bibinfo {title} {Single-molecule nucleic acid
  interactions monitored on a label-free microcavity biosensor platform},}\
  }\href@noop {} {\bibfield  {journal} {\bibinfo  {journal} {Nature
  nanotechnology}\ }\textbf {\bibinfo {volume} {9}},\ \bibinfo {pages}
  {933--939} (\bibinfo {year} {2014})}\BibitemShut {NoStop}%
\bibitem [{\citenamefont {Baaske}\ \emph {et~al.}(2022)\citenamefont {Baaske},
  \citenamefont {Asgari}, \citenamefont {Punj},\ and\ \citenamefont
  {Orrit}}]{Baaske2022}%
  \BibitemOpen
  \bibfield  {author} {\bibinfo {author} {\bibfnamefont {M.~D.}\ \bibnamefont
  {Baaske}}, \bibinfo {author} {\bibfnamefont {N.}~\bibnamefont {Asgari}},
  \bibinfo {author} {\bibfnamefont {D.}~\bibnamefont {Punj}}, \ and\ \bibinfo
  {author} {\bibfnamefont {M.}~\bibnamefont {Orrit}},\ }\bibfield  {title}
  {\enquote {\bibinfo {title} {Nanosecond time scale transient optoplasmonic
  detection of single proteins},}\ }\href {\doibase 10.1126/sciadv.abl5576}
  {\bibfield  {journal} {\bibinfo  {journal} {Science Advances}\ }\textbf
  {\bibinfo {volume} {8}} (\bibinfo {year} {2022}),\
  10.1126/sciadv.abl5576}\BibitemShut {NoStop}%
\bibitem [{\citenamefont {Lin}\ \emph {et~al.}(2021)\citenamefont {Lin},
  \citenamefont {Vermaas}, \citenamefont {Yan}, \citenamefont {de~Jong},\ and\
  \citenamefont {Prins}}]{lin2021click}%
  \BibitemOpen
  \bibfield  {author} {\bibinfo {author} {\bibfnamefont {Y.-T.}\ \bibnamefont
  {Lin}}, \bibinfo {author} {\bibfnamefont {R.}~\bibnamefont {Vermaas}},
  \bibinfo {author} {\bibfnamefont {J.}~\bibnamefont {Yan}}, \bibinfo {author}
  {\bibfnamefont {A.~M.}\ \bibnamefont {de~Jong}}, \ and\ \bibinfo {author}
  {\bibfnamefont {M.~W.}\ \bibnamefont {Prins}},\ }\bibfield  {title} {\enquote
  {\bibinfo {title} {Click-coupling to electrostatically grafted polymers
  greatly improves the stability of a continuous monitoring sensor with
  single-molecule resolution},}\ }\href@noop {} {\bibfield  {journal} {\bibinfo
   {journal} {ACS sensors}\ }\textbf {\bibinfo {volume} {6}},\ \bibinfo {pages}
  {1980--1986} (\bibinfo {year} {2021})}\BibitemShut {NoStop}%
\bibitem [{\citenamefont {Seth}\ \emph {et~al.}(2022)\citenamefont {Seth},
  \citenamefont {Mittal}, \citenamefont {Luan}, \citenamefont {Kolla},
  \citenamefont {Mazer}, \citenamefont {Joshi}, \citenamefont {Gupta},
  \citenamefont {Rathi}, \citenamefont {Wang}, \citenamefont {Morrissey} \emph
  {et~al.}}]{seth2022high}%
  \BibitemOpen
  \bibfield  {author} {\bibinfo {author} {\bibfnamefont {A.}~\bibnamefont
  {Seth}}, \bibinfo {author} {\bibfnamefont {E.}~\bibnamefont {Mittal}},
  \bibinfo {author} {\bibfnamefont {J.}~\bibnamefont {Luan}}, \bibinfo {author}
  {\bibfnamefont {S.}~\bibnamefont {Kolla}}, \bibinfo {author} {\bibfnamefont
  {M.~B.}\ \bibnamefont {Mazer}}, \bibinfo {author} {\bibfnamefont
  {H.}~\bibnamefont {Joshi}}, \bibinfo {author} {\bibfnamefont
  {R.}~\bibnamefont {Gupta}}, \bibinfo {author} {\bibfnamefont
  {P.}~\bibnamefont {Rathi}}, \bibinfo {author} {\bibfnamefont
  {Z.}~\bibnamefont {Wang}}, \bibinfo {author} {\bibfnamefont {J.~J.}\
  \bibnamefont {Morrissey}},  \emph {et~al.},\ }\bibfield  {title} {\enquote
  {\bibinfo {title} {High-resolution imaging of protein secretion at the
  single-cell level using plasmon-enhanced fluorodot assay},}\ }\href@noop {}
  {\bibfield  {journal} {\bibinfo  {journal} {Cell reports methods}\ }\textbf
  {\bibinfo {volume} {2}} (\bibinfo {year} {2022})}\BibitemShut {NoStop}%
\bibitem [{\citenamefont {Keng}, \citenamefont {Tan},\ and\ \citenamefont
  {Arnold}(2014)}]{Keng2014}%
  \BibitemOpen
  \bibfield  {author} {\bibinfo {author} {\bibfnamefont {D.}~\bibnamefont
  {Keng}}, \bibinfo {author} {\bibfnamefont {X.}~\bibnamefont {Tan}}, \ and\
  \bibinfo {author} {\bibfnamefont {S.}~\bibnamefont {Arnold}},\ }\bibfield
  {title} {\enquote {\bibinfo {title} {Whispering gallery micro-global
  positioning system for nanoparticle sizing in real time},}\ }\href {\doibase
  10.1063/1.4893762} {\bibfield  {journal} {\bibinfo  {journal} {Applied
  Physics Letters}\ }\textbf {\bibinfo {volume} {105}} (\bibinfo {year}
  {2014}),\ 10.1063/1.4893762}\BibitemShut {NoStop}%
\bibitem [{\citenamefont {Novotny}\ and\ \citenamefont
  {Hecht}(2012)}]{novotny2012principles}%
  \BibitemOpen
  \bibfield  {author} {\bibinfo {author} {\bibfnamefont {L.}~\bibnamefont
  {Novotny}}\ and\ \bibinfo {author} {\bibfnamefont {B.}~\bibnamefont
  {Hecht}},\ }\href@noop {} {\emph {\bibinfo {title} {Principles of
  nano-optics}}}\ (\bibinfo  {publisher} {Cambridge university press},\
  \bibinfo {year} {2012})\BibitemShut {NoStop}%
\bibitem [{\citenamefont {Eerqing}\ \emph {et~al.}(2023)\citenamefont
  {Eerqing}, \citenamefont {Wu}, \citenamefont {Subramanian}, \citenamefont
  {Vincent},\ and\ \citenamefont {Vollmer}}]{eerqing2023anomalous}%
  \BibitemOpen
  \bibfield  {author} {\bibinfo {author} {\bibfnamefont {N.}~\bibnamefont
  {Eerqing}}, \bibinfo {author} {\bibfnamefont {H.-Y.}\ \bibnamefont {Wu}},
  \bibinfo {author} {\bibfnamefont {S.}~\bibnamefont {Subramanian}}, \bibinfo
  {author} {\bibfnamefont {S.}~\bibnamefont {Vincent}}, \ and\ \bibinfo
  {author} {\bibfnamefont {F.}~\bibnamefont {Vollmer}},\ }\bibfield  {title}
  {\enquote {\bibinfo {title} {Anomalous dna hybridisation kinetics on gold
  nanorods revealed via a dual single-molecule imaging and optoplasmonic
  sensing platform},}\ }\href@noop {} {\bibfield  {journal} {\bibinfo
  {journal} {Nanoscale Horizons}\ } (\bibinfo {year} {2023})}\BibitemShut
  {NoStop}%
\bibitem [{\citenamefont {Vollmer}\ and\ \citenamefont
  {Arnold}(2008)}]{vollmer2008whispering}%
  \BibitemOpen
  \bibfield  {author} {\bibinfo {author} {\bibfnamefont {F.}~\bibnamefont
  {Vollmer}}\ and\ \bibinfo {author} {\bibfnamefont {S.}~\bibnamefont
  {Arnold}},\ }\bibfield  {title} {\enquote {\bibinfo {title}
  {Whispering-gallery-mode biosensing: label-free detection down to single
  molecules},}\ }\href@noop {} {\bibfield  {journal} {\bibinfo  {journal}
  {Nature methods}\ }\textbf {\bibinfo {volume} {5}},\ \bibinfo {pages}
  {591--596} (\bibinfo {year} {2008})}\BibitemShut {NoStop}%
\bibitem [{\citenamefont {Houghton}\ \emph {et~al.}(2024)\citenamefont
  {Houghton}, \citenamefont {Toropov}, \citenamefont {Yu}, \citenamefont
  {Bagby},\ and\ \citenamefont {Vollmer}}]{houghton2024single}%
  \BibitemOpen
  \bibfield  {author} {\bibinfo {author} {\bibfnamefont {M.~C.}\ \bibnamefont
  {Houghton}}, \bibinfo {author} {\bibfnamefont {N.~A.}\ \bibnamefont
  {Toropov}}, \bibinfo {author} {\bibfnamefont {D.}~\bibnamefont {Yu}},
  \bibinfo {author} {\bibfnamefont {S.}~\bibnamefont {Bagby}}, \ and\ \bibinfo
  {author} {\bibfnamefont {F.}~\bibnamefont {Vollmer}},\ }\bibfield  {title}
  {\enquote {\bibinfo {title} {Single molecule thermodynamic penalties applied
  to enzymes by whispering gallery mode biosensors},}\ }\href@noop {}
  {\bibfield  {journal} {\bibinfo  {journal} {Advanced Science}\ }\textbf
  {\bibinfo {volume} {11}},\ \bibinfo {pages} {2403195} (\bibinfo {year}
  {2024})}\BibitemShut {NoStop}%
\bibitem [{\citenamefont {Rothemund}(2006)}]{rothemund2006folding}%
  \BibitemOpen
  \bibfield  {author} {\bibinfo {author} {\bibfnamefont {P.~W.}\ \bibnamefont
  {Rothemund}},\ }\bibfield  {title} {\enquote {\bibinfo {title} {Folding dna
  to create nanoscale shapes and patterns},}\ }\href@noop {} {\bibfield
  {journal} {\bibinfo  {journal} {Nature}\ }\textbf {\bibinfo {volume} {440}},\
  \bibinfo {pages} {297--302} (\bibinfo {year} {2006})}\BibitemShut {NoStop}%
\bibitem [{\citenamefont {J{\"a}ger}, \citenamefont {Nir},\ and\ \citenamefont
  {Weiss}(2006)}]{jager2006site}%
  \BibitemOpen
  \bibfield  {author} {\bibinfo {author} {\bibfnamefont {M.}~\bibnamefont
  {J{\"a}ger}}, \bibinfo {author} {\bibfnamefont {E.}~\bibnamefont {Nir}}, \
  and\ \bibinfo {author} {\bibfnamefont {S.}~\bibnamefont {Weiss}},\ }\bibfield
   {title} {\enquote {\bibinfo {title} {Site-specific labeling of proteins for
  single-molecule fret by combining chemical and enzymatic modification},}\
  }\href@noop {} {\bibfield  {journal} {\bibinfo  {journal} {Protein Science}\
  }\textbf {\bibinfo {volume} {15}},\ \bibinfo {pages} {640--646} (\bibinfo
  {year} {2006})}\BibitemShut {NoStop}%
\end{thebibliography}%

\end{document}